\documentclass[aps,showpacs]{revtex4}
\usepackage{amsmath}
\usepackage{graphicx}
\usepackage{amssymb}
\usepackage{hyperref}
\usepackage{subfig}

\begin{document}

\title{Connecting boundary and interior \textemdash{} {}``Gauss's law''
for graphs}

\author{Ching King \surname{Chan}}

\author{Kwok Yip \surname{Szeto}}\email{phszeto@ust.hk}

\affiliation{Deparatment of Physics, The Hong Kong University of Science and Technology, Clear Water Bay, Kowloon, Hong Kong, China}

\begin{abstract}
The Gauss's law, in an abstract sense, is a theorem that relates quantities
on the boundary (flux) to the interior (charge) of a surface. An identity
for soap froths were proved with the same boundary--interior relation.
In this article, we try to construct a definition of flux for other
graphs, such that a similar boundary--interior relation can be satisfied. 
\end{abstract}
\maketitle

\section{Introduction}
The introduction of random lattice for the description of space-time continuum
 has provided a new perspective to the foundation of physics
\cite{Christ:1982fk,Christ:1982uq,Sahlmann:2010kx}.
The random lattice consists of points distributed at random, with nearby sites
 connected so as to form the edges of non-overlapping simplices which fill the
 entire system.
Lorentz invariance can be maintained by proper averaging over these random
 lattices.

Rather than investigating the consequence of this radical approach to physics,
 we like to look at the random lattice more closely in the perspective of
 networks.
If one consider the simplices formed by nodes and links in the random lattice as
 a complex network, we can ask about its general properties, such as degree
 distribution, clustering coefficients, etc of the random lattice and see if
 these properties of the network have important relation to the field theories
 that are build on such discretization of space-time.

For now, we still want to narrow down the investigation to a more mundane
 question, which concerns the relevance of the random lattice in the
 discretization of classical electrodynamics.
As Gauss's law is fundamental to classical electrodynamics, we like to see if a
 generalization of this law can be formulated for networks, and if so, what kind
 of networks will obey Gauss's law.
One immediate example is the trivalent cellular network formed by soap froth.
This cellular network has a simple property relating the boundary to the
 interior, through a quantity called topological charge.
We like to generalize this observation in soap froth to complex networks.

\section{Theory}
\subsection{Inspiration\label{sub:Inspiration}}

\begin{figure}
\subfloat[\label{fig:Soap-froth}Soap froth. The dashed lines show
the dual-graph representation. The cluster consists of a pentagon
and a hexagon, so the total charge is $Q=(6-5)+(6-6)=1$. Its boundary
has 7 convex vertices (labeled {}``$+$'') and 2 concave vertices
(labeled {}``$-$''), so the total flux is $\Phi=6-7+2=1$, and
we see that $Q=\Phi$]{\includegraphics[width=0.375\textwidth]{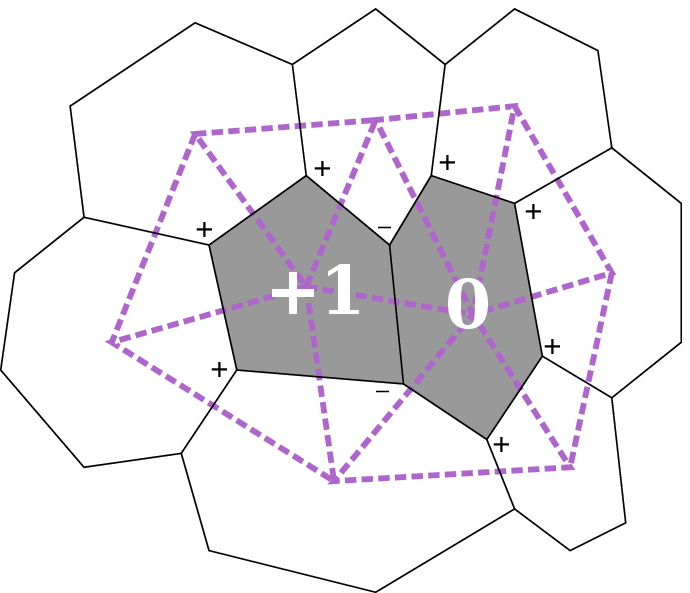}}
\qquad
\subfloat[\label{fig:Trees:}Trees. The enclosed nodes have degrees
5, 1 and 4 respectively, so the total charge (as indicated by the
numbers) is $(2-5)+(2-1)+(2-4)=-4$. There are 6 edges crossing the
boundary, so the flux is $2-6=-4$ too.]{\includegraphics[width=0.375\textwidth]{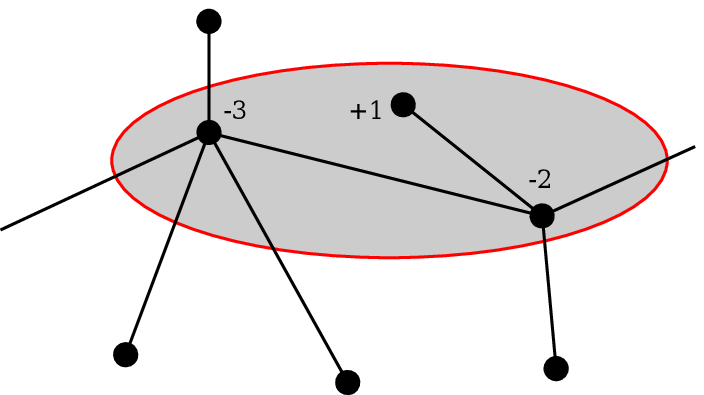}}
\caption{Examples of how {}``Gauss's law'' is satisfied for (a) soap froths
and (b) trees.}
\end{figure}

Bubbles in 2D soap froth carry a property known as topological charge\cite{Stavans:1989uq},
6 minus the number of sides of the bubble. With a simple cluster of
bubbles, it was shown that the total charge of the cluster is related
to the vertices on the cluster's edge\cite{Aste:1996fk}, which is
a simple application of Euler's formula $V-E+F=2$. The soap froth
is in fact an infinite trivalent graph, with the bubbles as nodes,
and contacting faces as edges. In this dual-graph representation,
we define the charge $q_{v}$ of a node $v$ as $6-\deg v$, and the
flux $\Phi(S)$ of a connected cluster of nodes $S$ as the number
of triangles with exactly 1 node outside $S$, minus the number of
triangles with exactly 1 node inside $S$, plus 6. Then we find that
(Figure \ref{fig:Soap-froth})\begin{equation}
\Phi(S)=\sum_{v\in S}q_{v}.\label{eq:gauss}\end{equation}
This can be considered as a topological analog of Gauss's law $\Phi(S)=\oint_{S}\mathbf{E}\cdot d\mathbf{S}=Q/\epsilon_{0}$
as the $\Phi$ term only related to quantities on the boundary \textemdash{}
objects that span across the interior and exterior of $S$ \textemdash{}
and the $Q$ term is an accumulation of quantities inside $S$.

The boundary--interior relation can be found on undirected trees (i.e.\ 
acyclic graphs) as well. If we define the charge of a tree vertex
$v$ as $q_{v}=2-\deg v$, and the flux $\Phi$ of a cluster as 2
minus the number of edges crossing the boundary, then Equation \ref{eq:gauss}
will also be satisfied (Figure \ref{fig:Trees:}).

From these examples, we conjecture that with a suitable definition
of $q_{v}$ and $\Phi(S)$, a corresponding Equation \ref{eq:gauss}
can apply to some classes of graphs. We would like to formulate
an expression for $\Phi$ and $q_{v}$, and find out some examples
and counter-examples.

\subsection{Flux}

In the following, we suppose $G=(V,E)$ is a simple undirected graph
and is connected.

In both the trivalent graph and tree cases, the flux is defined in
terms of count of objects that crosses the boundary. Therefore, we
can define the flux of a set of nodes $S$ inducing a connected
subgraph as a count of arbitrary subgraphs across the boundary:
\[
\Phi''(S)=A_{0}\left(S\right)+\sum_{\substack{H\subseteq G\\
H\cap S\ne\varnothing\\
H\cap\left(V\setminus S\right)\ne\varnothing}
}A_{H,\left|H\cap S\right|},\]
where $A_{0}(S)$ is an arbitrary real function, and $A_{H,\left|H\cap S\right|}$
is a real constant depending on $H$ and $\left|H\cap S\right|$.
We call $\Phi''$ the \emph{partial flux} of $S$. Like a physical
system, this artificial flux defined here should also satisfy some
preconditions. The most important one is the \emph{additivity of flux}:\begin{equation}
\Phi''\left(S_{1}\cup S_{2}\right)=\Phi''\left(S_{1}\right)+\Phi''\left(S_{2}\right)\label{eq:additive}\end{equation}
when $S_{1}\cap S_{2}=\varnothing$ and $S_{1}\cup S_{2}$ induces
a connected subgraph. This gives the definition of partial charge
in terms of partial flux when we enforce Equation \ref{eq:gauss}:\[
q_{v}''=\Phi''\left(\left\{ v\right\} \right)\]
because \[
\Phi''\left(S\right)=\Phi''\left(\bigcup_{v\in S}\left\{ v\right\} \right)=\sum_{v\in S}\Phi''\left(\left\{ v\right\} \right)=\sum_{v\in S}q_{v}''.\]

Looking at the definition of fluxes of soap froths and trees, we find
that there is already a constant term {}``6'' and {}``2'', so
for simplicity we assume $A_{0}(S)$ is just a constant. The subgraphs
$H$ are usually chosen with special properties, such as cliques or
cycles. For example, if we choose $H$ to be all complete subgraphs,
then, applying distributive law, the partial flux can
be rewritten as \begin{equation}
\Phi''(S)=A_{0}+\sum_{n=3}^{\infty}\sum_{l=1}^{\lceil n/2\rceil-1}A_{nl}F_{nl}(S),\label{eq:clique-flux}\end{equation}
where $F_{nl}(S)$ is the count of complete $n$-subgraphs having
$l$ nodes inside $S$, and $A_{nl}$ are coefficients derived from
the graph's property. Since for a graph $G$ with $N$ nodes,
the highest clique number is $N$, so the infinite sum will stop when
$n=N$. Thus, there will be at most $\sum_{n=3}^{N}\left(\lceil n/2\rceil-1\right)\sim N^{2}/4$
coefficients.

In the above we assumed both $S$ have exactly 1 connected component.
When $S$ is composed of many connected components $S_{i}$, we impose
Equation \ref{eq:additive} and define the partial flux for disconnected
subgraphs\[
\Phi'(S)=\sum_{i}\Phi''(S_{i}).\]

Now consider the flux of the whole graph. This set has no boundary,
so $\Phi'\left(V\right)=A_{0}$. In particular, $\Phi'(V)$ has no
contribution from $A_{H,|H\cap S|}$. Suppose we split $V$ into two
subset of nodes $S$ and $V\setminus S$ inducing two subgraphs. Many
subgraphs $H$ in the interior of $V$ will now reside on the boundary
of $S$ and $V\setminus S$. Enforcing the additivity of flux, these
contributions should cancel each other. We could antisymmetrize the
partial flux to give the \emph{complete flux} that expresses this:\[
\Phi(S)=A_{0}+\Phi'(S)-\Phi'\left(V\setminus S\right).\]
The extra $A_{0}$ is needed to maintain the invariance of $\Phi(V)$
for whatever partition $\{S,V\setminus S\}$ we choose. The complete
charge is thus similarly defined\[
q_{v}=\Phi\left(\left\{ v\right\} \right)\]

A flux that is not identically zero may not be sensibly definable
for every graph.

\section{Examples}

\subsection{Undirected Trees}

Trees has partial flux\[
\Phi''_{\text{tree}}(S)=1.\]
This is equivalent to the charge defined in Section \ref{sub:Inspiration}.
When a node $v$ has degree $k$, the subgraph $V\setminus\{v\}$
will have $k$ connected components. Therefore,\begin{align*}
\Phi\left(\left\{ v\right\} \right) & =1+\Phi'\left(\left\{ v\right\} \right)-\Phi'\left(V\setminus\left\{ v\right\} \right)\\
 & =1+1-\sum_{i=1}^{k}1\\
 & =2-k.\end{align*}
The tree is the simplest family of graphs that have a well-defined
flux, which the partial flux is just a constant. 

This is equivalent to the property that every connected
subgraph of a tree is still a tree, and a tree with $N$ nodes has
exactly $N-1$ edges. Suppose we pick any connected set of nodes $S$
of the tree $G$, then the induced subtree it has $\left|S\right|$
nodes and $\left|S\right|-1$ edges. Using the handshaking lemma,
i.e.\ the total degree of a graph $G'$ is twice the number of edges
$\sum_{v\in G'}k_{v}=2\left(\#\text{edges of }G'\right)$, we have
\begin{align*}
\sum_{v\in S}k_{v} & =2\left(\#\text{edges within }S\right)+\left(\#\text{edges out of }S\right)\\
 & =2\left|S\right|-2+\left(\#\text{edges out of }S\right)\\
\sum_{v\in S}\underbrace{\left(2-k_{v}\right)}_{\Phi\left(\left\{ v\right\} \right)} & =\underbrace{2-\left(\#\text{edges out of }S\right)}_{\Phi\left(S\right)}.\end{align*}
This proves our definition of flux is correct.

\subsection{Triangular Graphs}

\begin{figure}
\includegraphics[width=0.25\textwidth]{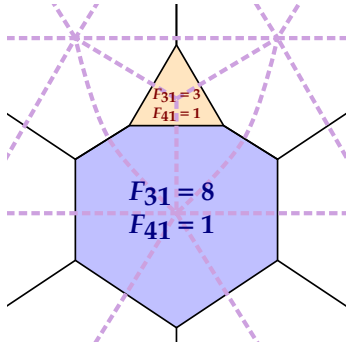}
\caption{\label{fig:charge-defect}There is no well-defined flux when a defect
appears on the boundary. As shown above, the heptagonal
node is part of 8 different triangles $K_{3}$ and 1 tetrahedron $K_{4}$
($F_{31}=8,F_{41}=1$) in the dual graph, which has an overcount of
triangle connecting the three heptagonal nodes. However, the triangular
node is part of 3 different triangles $K_{3}$ and 1 tetrahedron $K_{4}$
($F_{31}=3,F_{41}=1$) which does not have any overcount.}
\end{figure}

Triangular graphs are graphs which the only faces are triangles. The
2D soap froth is of this kind. The partial flux could be defined as\[
\Phi''_{\text{triangular}}(S)=6-F_{31}(S),\]
where $F_{31}(S)$ is the same as the definition in Equation \ref{eq:clique-flux}.
Considering the dual graph of triangular graph, $F_{31}(S)$ is the
same as $V^{+}(S)$ in Reference \cite{Aste:1996fk}, so this definition
of partial flux also makes sense \textemdash{} as long as $S$ does
not contain a $K_{4}$ (4-clique, a.k.a. tetrahedron)
on its boundary, i.e.\ a defect (triangular bubble) in the soap
froth. 

The error when including a defect is due to overcounting of triangles
\textemdash{} there should be only 3 triangles in $K_{4}$ in the
planar graph, but when the embedded space is not considered, 4 triangles
would be counted. When the defect causes an overcount, this leads
to an extra term\[
\Phi''_{\text{triangular}}(S)=6-F_{31}(S)+F_{41}(S),\]
but in general defect doesn't always overcount, e.g.\ the charge of
a defect (Figure \ref{fig:charge-defect}). Therefore, the triangular
graph's flux is well-defined as long as the defect does not appear
on the boundary.

Since the flux of triangular graph is essentially derived from Euler's
formula, it should be generalizable to higher-dimensional simplicial
graphs (soap froths).

\section{Discussion}

\begin{figure*}
\subfloat[A graph that has no well-defined flux.]{
\includegraphics[width=0.25\textwidth]{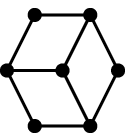}
}
\qquad
\subfloat[Flux vs charge]{
\includegraphics[width=0.5\textwidth]{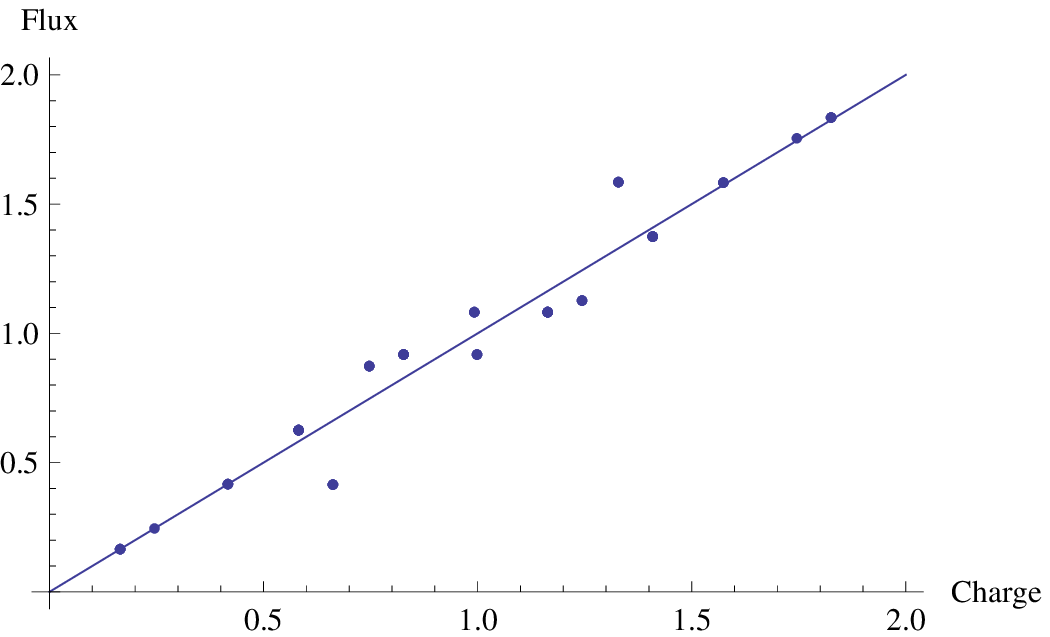}
}\caption{\label{fig:non-gauss}If we plot the flux vs total charge for all connected
subgraphs of the graph in (a), the values do not fall in a straight
line completely. Linear regression gives $\Phi''(S)=1-0.0857F_{41}^{\circ}(S)-0.1659F_{61}^{\circ}(S)-0.0410F_{62}^{\circ}(S)$
with $r^{2}=0.9987$.}
\end{figure*}

\begin{table*}
\begin{tabular}{cc}
\hline 
Graph & Partial flux $\Phi''(S)$\tabularnewline
\hline
\hline 
All of \includegraphics{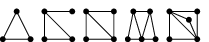} & $3-F_{31}$\tabularnewline
\hline 
Hamiltonian graphs\cite{Diestel:2006fk} with 4 nodes & $4-2F_{41}^{\circ}$\tabularnewline
\hline 
Hamiltonian graphs with 5 nodes & $5-3F_{51}^{\circ}-F_{52}^{\circ}$\tabularnewline
\hline 
Hamiltonian graphs with 6 nodes & $6-4F_{61}^{\circ}-2F_{62}^{\circ}$\tabularnewline
\hline 
\includegraphics{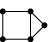} & $6-2F_{31}^{\circ}-3F_{41}^{\circ}$\tabularnewline
\hline 
\includegraphics{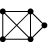} & $6-2F_{31}-3F_{41}$\tabularnewline
\hline 
Tetrahedron, Octahedron, Icosahedron & $6-F_{31}$\tabularnewline
\hline 
Cube and \includegraphics{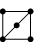} & $4-F_{41}^{\circ}$\tabularnewline
\hline 
Dodecahedron & $10-3F_{51}^{\circ}-F_{52}^{\circ}$\tabularnewline
\hline 
\includegraphics{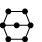} & $30-9F_{51}^{\circ}-3F_{52}^{\circ}-10F_{61}^{\circ}-5F_{62}^{\circ}$\tabularnewline
\hline
\end{tabular}\caption{\label{tab:well-def-flux}Some small graphs that have well-defined
fluxes.}
\end{table*}

We see that there isn't a general rule on which types of subgraphs
$H$ should be chosen, so it is hard to deny the existence of a definition
of nonzero partial flux for a particular graph $G=(V,E)$. However,
we could still experimentally prove or reject some possibilities of
flux, based on the equality of flux and total charge (Equation \ref{eq:gauss}). 

Assume the partial flux is defined with Equation \ref{eq:clique-flux},
where $A_{0}$ and $A_{nl}$ are assumed to be unknown. Given an arbitrary
induced subgraph $S$, we could compute $F_{nl}(S)$, and thus $\Phi(S)$
and $Q(S)=\sum_{v\in S}q_{v}$. Both expressions are linear combinations
$A_{0}$ and $A_{nl}$. If the flux is well-defined, $\Phi(S)$ and
$Q(S)$ must be equal, i.e.
\[\Phi(S)-Q(S)=0,\]
and this yields one linear equation involving $A_{0}$ and $A_{nl}$.
There are $2^{|V|}$ possible subgraphs (thus equations), while the
number of coefficients $A_{nl}$ is at most $\left|V\right|^{2}/4$.
If this overspecified problem can be solved with a nontrivial solution,
i.e.\ the matrix defined by this homogeneous system of linear equations
have nonzero nullity, then the solution gives the definition of partial
flux. In practice, $2^{|V|}$ is too big even for a relatively small
graph, so we may randomly pick $X\gg\left|V\right|^{2}/4$ subgraphs
to check. We could also use linear regression to find a best fit,
but this may not report the expected result.

Using the above method, we could find that highly symmetrical
graphs such as the cube graph and dodecahedron also have well-defined
fluxes
\begin{align*}
\Phi''_{\text{cube}}(S) & =4-F_{41}^{\circ}(S)\\
\Phi''_{\text{dodecahedron}}(S) & =10-3F_{51}^{\circ}(S)-F_{52}^{\circ}(S)\end{align*}
where $F_{nl}^{\circ}(S)$ is similar to $F_{nl}(S)$, except
that cycles are counted instead of cliques. Table \ref{tab:well-def-flux}
shows other linear regression results. However, many other graphs
do not have a well-defined flux either when counting cliques or simple
cycles. A simple counter-example is shown in Figure \ref{fig:non-gauss},
where Equation \ref{eq:clique-flux} is not well-defined. We have
seen a number of positive and negative examples already, so we question,
what kind of graphs will have a well-defined flux? In particular,
do physical graphs defined spatially like the soap froth all have
a mostly well-defined flux? If true, this would place an strong restriction
on what graphs would be allowed.

Another food for thought is what family of graphs can be generated
given a definition of partial flux, e.g.\ would only trees satisfy
$\Phi''(S)=1$? The partial flux could give an interesting classification
of graphs.
\begin{acknowledgments}
K.\ Y.\ Szeto acknowledges the support of CERG602507 grant.

\end{acknowledgments}
\bibliography{bib}

\begin{thebibliography}{6}
\expandafter\ifx\csname natexlab\endcsname\relax\def\natexlab#1{#1}\fi
\expandafter\ifx\csname bibnamefont\endcsname\relax
  \def\bibnamefont#1{#1}\fi
\expandafter\ifx\csname bibfnamefont\endcsname\relax
  \def\bibfnamefont#1{#1}\fi
\expandafter\ifx\csname citenamefont\endcsname\relax
  \def\citenamefont#1{#1}\fi
\expandafter\ifx\csname url\endcsname\relax
  \def\url#1{\texttt{#1}}\fi
\expandafter\ifx\csname urlprefix\endcsname\relax\def\urlprefix{URL }\fi
\providecommand{\bibinfo}[2]{#2}
\providecommand{\eprint}[2][]{\url{#2}}

\bibitem[{\citenamefont{Christ et~al.}(1982{\natexlab{a}})\citenamefont{Christ,
  Friedberg, and Lee}}]{Christ:1982fk}
\bibinfo{author}{\bibfnamefont{N.~H.} \bibnamefont{Christ}},
  \bibinfo{author}{\bibfnamefont{R.}~\bibnamefont{Friedberg}},
  \bibnamefont{and} \bibinfo{author}{\bibfnamefont{T.~D.} \bibnamefont{Lee}},
  \bibinfo{journal}{Nucl. Phys. B} \textbf{\bibinfo{volume}{202}},
  \bibinfo{pages}{89} (\bibinfo{year}{1982}{\natexlab{a}}).

\bibitem[{\citenamefont{Christ et~al.}(1982{\natexlab{b}})\citenamefont{Christ,
  Friedberg, and Lee}}]{Christ:1982uq}
\bibinfo{author}{\bibfnamefont{N.~H.} \bibnamefont{Christ}},
  \bibinfo{author}{\bibfnamefont{R.}~\bibnamefont{Friedberg}},
  \bibnamefont{and} \bibinfo{author}{\bibfnamefont{T.~D.} \bibnamefont{Lee}},
  \bibinfo{journal}{Nucl. Phys. B} \textbf{\bibinfo{volume}{210}},
  \bibinfo{pages}{337} (\bibinfo{year}{1982}{\natexlab{b}}).

\bibitem[{\citenamefont{Sahlmann}(2010)}]{Sahlmann:2010kx}
\bibinfo{author}{\bibfnamefont{H.}~\bibnamefont{Sahlmann}},
  \bibinfo{journal}{Phys. Rev. D} \textbf{\bibinfo{volume}{82}},
  \bibinfo{pages}{064018} (\bibinfo{year}{2010}), \eprint{0911.4180}.

\bibitem[{\citenamefont{Stavans and Glazier}(1989)}]{Stavans:1989uq}
\bibinfo{author}{\bibfnamefont{J.}~\bibnamefont{Stavans}} \bibnamefont{and}
  \bibinfo{author}{\bibfnamefont{J.~A.} \bibnamefont{Glazier}},
  \bibinfo{journal}{Phys. Rev. Lett.} \textbf{\bibinfo{volume}{62}},
  \bibinfo{pages}{1318} (\bibinfo{year}{1989}).

\bibitem[{\citenamefont{Aste et~al.}(1996)\citenamefont{Aste, Szeto, and
  Tam}}]{Aste:1996fk}
\bibinfo{author}{\bibfnamefont{T.}~\bibnamefont{Aste}},
  \bibinfo{author}{\bibfnamefont{K.~Y.} \bibnamefont{Szeto}}, \bibnamefont{and}
  \bibinfo{author}{\bibfnamefont{W.~Y.} \bibnamefont{Tam}},
  \bibinfo{journal}{Phys. Rev. E} \textbf{\bibinfo{volume}{54}},
  \bibinfo{pages}{5482} (\bibinfo{year}{1996}).

\bibitem[{\citenamefont{Diestel}(2006)}]{Diestel:2006fk}
\bibinfo{author}{\bibfnamefont{R.}~\bibnamefont{Diestel}},
  \emph{\bibinfo{title}{Graph theory}} (\bibinfo{publisher}{Springer},
  \bibinfo{year}{2006}), chap.~\bibinfo{chapter}{10}, p. \bibinfo{pages}{275},
  \bibinfo{edition}{3rd} ed.

\end{thebibliography}

\end{document}